# Analysis of the Reliability of a Biofuel Production Plant from Waste Cooking Oil

Ivan Nekrasov[1, a)], Aleksandr Zagulyaev[1, b)], Vladimir Bukhtoyarov[1, 3, c)], Svetlana Eremeeva[1, d)], Elena Filyushina[1, 2, e)], Aleksey Gorodov[1, 2, f)] and Natalia Shepeta[1, 2, g)]

[1]*Siberian Federal University, 79, Svobodny Av., Krasnoyarsk, 660041, Russia*
[2]*Reshetnev Siberian State University of Science and Technology, 31, Krasnoyarskiy rabochiy pr., Krasnoyarsk, Russia*
[3]*Bauman Moscow State Technical University, 5, 2-nd Baumanskaya, Moscow, 105005, Russia*

a) Corresponding author: nekrasov-is@ya.ru
b) zagulyaev.aleksandr@mail.ru
c) vbukhtoyarov@sfu-kras.ru
d) seremeeva@sfu-kras.ru
e) efilyushina@sfu-kras.ru
f) glexx84@mail.ru
g) nshepeta@sfu-kras.ru

**Abstract.** The article considers the issue of increasing the structural reliability of a biofuel production plant. A review of the existing basic technological schemes of the biofuel production plant has been carried out. The main structural elements are determined and a functional diagram is constructed. Processed cooking oil was chosen as the input raw material. A structural analysis of the reliability of each element and the entire system as a whole was carried out. The least reliable elements are determined, options for improving the overall reliability of the installation are proposed.

## INTRODUCTION

Recently, due to the tightening of environmental requirements and the increase in the price of petroleum products, the demand for the use of biofuels and biofuel compositions has increased [1-2]. Biofuel composition - a mixture of mineral raw materials with fuel of vegetable or animal origin or organic industrial waste. Used cooking oil or various oils of vegetable crops can act as a source of raw materials. For example: camelina oil, soybean oil, corn oil, rapeseed oil [3-4].

The production process takes place with or without a catalyst (under supercritical conditions) through an etherification reaction. Several variants of the basic technological schemes of a biofuel production plant are considered, and in general they contain several main blocks: a feedstock supply unit, a reactor unit, a separation unit, and a rectification unit [5–6].

The papers [7-8] analyze the reliability of a biofuel production plant from the point of view of the human factor and hidden failures in production. We will carry out the analysis based on the theory of reliability.

The purpose of the article is to conduct a structural analysis of the reliability of a biofuel production plant, which uses waste cooking oil as a raw material.

## METHODS

Reliability is the property of an object to keep in time within the established limits the values of all parameters that characterize the ability to perform the required functions in given modes and conditions of use, maintenance, repairs, storage and transportation.

The reliability of an object over time is characterized by the value $R(t)$, whose values lie in the interval [0;1]. Since equipment tends to fail over time, their reliability will decrease, which means $R(t)$ is a decreasing function of $t: R'(t) < 0$. At the same time, we assume that at the initial moment of time, the reliability is equal to 1: $R(0) = 1$. In this article, we will use the exponential dependence of reliability on time: $R(t) = e^{-\lambda t}$, where $\lambda$ - failure rate. Failure rate is a value that characterizes the amount of equipment that has failed per unit of time. For each type of equipment, this is a kind of constant that will be used in calculations.

A system with a serial connection of elements is a system in which the failure of any element leads to the failure of the entire system (Fig. 1).

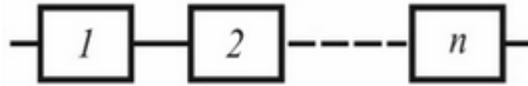

**FIGURE 1.** Serial connection of elements.

With such a connection of elements, the reliability of the entire system is calculated as the product of the reliability of individual elements of the system:

$$R = \prod_{i=1}^{n} R_i \qquad (1)$$

A system with parallel connection of elements is a system whose failure occurs only in the event of failure of all its elements (Fig. 2).

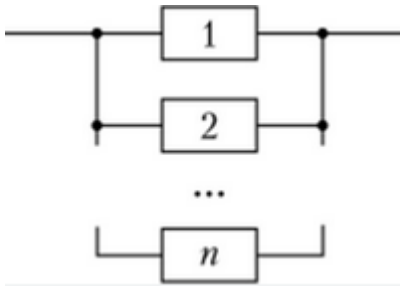

**FIGURE 2.** Parallel connection of elements

With such a connection of elements, the reliability of the entire system is calculated as follows:

$$R = 1 - \prod_{i=1}^{n}(1 - R_i) \qquad (2)$$

The technological scheme of the hydrocracking unit is shown in Fig. 3.

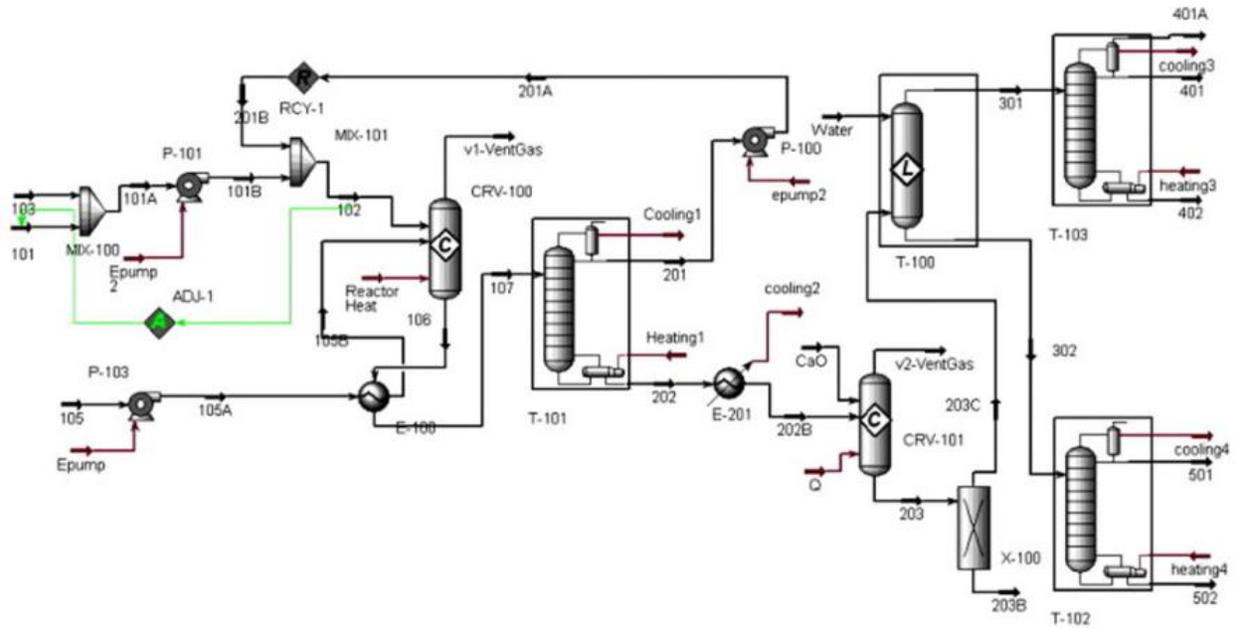

**FIGURE 3.** Technological scheme of the biofuel production unit.

## RESULTS

On the basis of the technological scheme, a functional diagram of the biofuel production unit was built (Fig. 4).

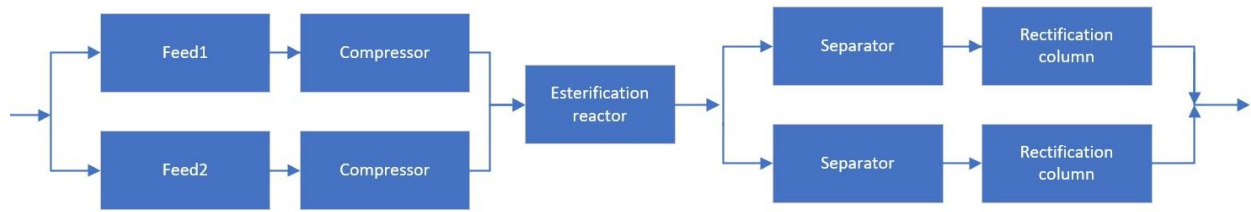

**FIGURE 4.** Functional diagram of a biofuel production plant.

Reliability of the feed system

$$R_1 = R_{11} R_{12} R_{13} \tag{3}$$

$$R_{13} = R_P R_V \tag{4}$$

where $R_{11}$ – reliability of the reactor fresh feed filter, $R_{12}$ - reliability of the reactor fresh feed filter, $R_V$ – reliability of the surge vessel supplying the reactor, $R_P$ – reliability of the reactor feed pump. The corresponding failure rates are given in Table 1:

**TABLE 1.** Failure rate of raw material supply equipment

| Title | Failure rate |
|---|---|
| Reactor Fresh Feed Filter | 0.000031 |
| Reactor Recycle Feed Filter | 0.000006 |
| Reactor Feed Surge Vessel | 0.000001 |
| Reactor Feed Pump | 0.000003 |

Taking into account that $R(t) = e^{-\lambda t}$ and a duration of 1000 hours, we get the reliability for the supply system $R_1 = 0.9598$.

To calculate the reliability of the remaining elements of the scheme, we turn to Table 2, which reflects the failure rate for the equipment used in the biofuel production unit:

**TABLE 2.** Biofuel production unit operating parameters

| Title | Failure rate |
|---|---|
| Heat exchanger | 0.00002 |
| Heater | 0.000037 |
| Reactor | 0.000007 |
| High pressure separator | 0.000013 |
| Low pressure separator | 0.00005 |
| Distillation column | 0.000001 |

For the whole scheme, we obtain the following reliability:

$$R = R_1 R_2^2 R_3 R_4 (1-(1-R_5)(1-R_6))(1-(1-R_7)^2) \tag{5}$$

where $R_2$ - heat exchanger reliability, $R_3$ - furnace reliability, $R_4$ - reactor reliability, $R_5$ - high pressure separator reliability, $R_6$ - low pressure separator reliability, $R_7$ - distillation column reliability. Considering the above failure rates, we will have $R_2 = 0.98$, $R_3 = 0.964$, $R_4 = 0.993$, $R_5 = 0.987$, $R_6 = 0.951$, $R_7 = 0.999$. And the overall reliability (for 1000 h) of the installation as a whole $R = 0.882$.

## DISCUSSION

Despite the fact that the reliability of the entire system turned out to be quite high, there is room for improvement in terms of reliability. The least reliable elements of the system are the furnace and the low-pressure separator. Redundancy of these elements of the system, as well as more frequent maintenance, can be a way to solve this problem.

## CONCLUSIONS

The article considered a method for assessing the reliability of equipment using the example of a biofuel production unit. The installation with its elements was presented in the form of a block diagram, according to which the reliability of the entire system was determined by the reliability of its components. The least reliable elements: an oven with

$R_3 = 0.964$ and $R_6 = 0.951$. And the reliability of the whole system $R = 0.882$, which means that the system will not fail with a probability of 0.118 during 1000 hours of operation.

## ACKNOWLEDGMENTS

This study is carried out under the state assignment under the project "Development of a set of scientific and technical solutions in the field of creating biofuels and optimal biofuel compositions that provide the possibility of transforming consumed types of energy carriers in accordance with energy efficiency trends, reducing the carbon footprint of products and using alternative fuels to fossil fuels" (FSRZ Contract -2021-0012) in the scientific laboratory of biofuel compositions of the Siberian Federal University, created as part of the activities of the Research and Educational Center "Yenisei Siberia".